\documentclass[reprint,
 amsmath,amssymb,
prb,
]{revtex4-2}
\usepackage{graphicx}
\usepackage[usenames,dvipsnames]{xcolor}
\usepackage{caption}
\usepackage{subcaption}
\usepackage{comment}
\usepackage{dcolumn}
\usepackage{bm}
\usepackage{physics}
\usepackage{xcolor}

\begin{document}

\preprint{APS/123-QED}

\title{Field-free anomalous junction and superconducting diode effect in   spin split superconductor/topological insulator junctions}

\author{T.H. Kokkeler}
\email{tim.kokkeler@dipc.org}
\affiliation{Donostia International Physics Center (DIPC), 20018 Donostia--San Sebasti\'an, Spain\\
}%
\affiliation{Interfaces and Correlated Electron Systems,
 Faculty of Science and Technology, University of Twente, Enschede, The Netherlands\\
}


\author{A.A. Golubov}
\affiliation{Interfaces and Correlated Electron Systems,
 Faculty of Science and Technology, University of Twente, Enschede, The Netherlands\\}%
 
 \author{F. S. Bergeret}
\affiliation{Centro de F\'isica de Materiales (CFM-MPC) Centro Mixto CSIC-UPV/EHU, E-20018 Donostia-San Sebasti\'an,  Spain\\
}%
\affiliation{Donostia International Physics Center (DIPC), 20018 Donostia--San Sebasti\'an, Spain}


\date{\today}

\begin{abstract}
We study the transport properties of a diffusive Josephson junction between two spin-split superconductors made of superconductor-ferromagnetic insulator bilayers (FIS)  on top of a 3D topological insulator (TI). We derive the corresponding Usadel equation describing the quasiclassical Green's functions in these systems and first solve the equation analytically in the weak-proximity case.   We demonstrate the appearance of an anomalous phase in the absence of an external magnetic field. 
We also explore non-reciprocal electronic transport. Specifically, we calculate the junction's diode efficiency $\eta$ by solving the Usadel equation.
We obtain a sizable diode effect even at zero applied magnetic field. We discuss how the diode efficiency $\eta$ depends on the different parameters and find a non-monotonic behavior of $\eta$ with temperature. 

\end{abstract}

\maketitle

\section{Introduction}
Recent advances attracting attention in superconductivity research are effects related to non-reciprocal charge transport \cite{zazunov2009anomalous,tanaka2009manipulation,liu2010anomalous,alidoust2013varphi,alidoust2020critical,alidoust2021cubic,szombati2016josephson,brunetti2013anomalous,campagnano2015spin,lu2015anomalous}, particularly the superconducting diode effect \cite{wakatsuki2017nonreciprocal,wakatsuki2018nonreciprocal,ando2020observation,pal2022josephson,baumgartner2022supercurrent,hou2022ubiquitous,lin2022zero,lyu2021superconducting,wu2021realization,silaev2014diode,yuan2022supercurrent,tanaka2022theory,pal2019quantized,kopasov2021geometry,margaris2010zero,chen2018asymmetric,he2022phenomenological}. A mesoscopic superconducting junction is called a superconducting diode if the critical current is different for opposite current directions. That is,  for a superconducting diode the minimum, $I_{-} = \min_{\phi}{I_{s}}(\phi)$,  and maximum, $I_{+} = \max_{\phi}{I_{s}}(\phi)$, of the current phase relation (CPR) are unequal in magnitude. If a current $I$ flows in such a junction, with  $\min(|I_{-}|,I_{+})<I<\max(|I_{-}|,I_{+})$, in one direction it is a supercurrent, whereas in the other direction the current dissipates. 
In a conventional Josephson junction the critical current is the same in both directions, the CPR has the following symmetry: $I(-\phi) = -I(\phi)$ \cite{golubov2004current}. This symmetry holds 
if either time reversal symmetry or inversion symmetry is present in the system \cite{silaev2017anomalous,zhang2021general}.

The superconducting diode effect can thus only be obtained if both time reversal symmetry and inversion symmetry are broken \cite{wakatsuki2017nonreciprocal,wakatsuki2018nonreciprocal}. 
Time reversal symmetry breaking can be achieved by  a  magnetic field. On the other hand inversion symmetry can be broken intrinsically, such as in  topological insulators \cite{hasan2010colloquium,dolcini2015topological,karabassov2022hybrid} or superconductors with Rashba spin orbit coupling \cite{reynoso2008anomalous,buzdin2008direct,
yokoyama2014anomalous,
konschelle2015theory,bergeret2015theory,daido2022intrinsic,ilic2021effect,jeon2022zero}.  Inversion symmetry can also be broken by using an asymmetric junction geometry \cite{souto2022josephson} or  asymmetry of  the device originated in the  fabrication \cite{hou2022ubiquitous}.

If both time-reversal symmetry and inversion-symmetry are broken, in general $I(-\phi) \neq -I(\phi)$ and thus possibly $I(\phi = 0)\neq 0$. Such junctions, for which the current vanishes at a nonzero phase difference,  are called $\phi_{0}$-junctions  \cite{buzdin2008direct}. In weak coupling Josephson junctions, the CPR is proportional to $\sin(\phi+\phi_0)$; hence, in this regime, one cannot observe the diode effect.  However, if higher harmonics contribute to the current, in general $I_{+}\neq|I_{-}|$, and the diode effect can be observed  \cite{baumgartner2022supercurrent,ando2020observation, lin2022zero,lyu2021superconducting}. 

One way of breaking  time-reversal symmetry  without an external magnetic field is to attach a ferromagnetic insulator (FI) to the superconductor. FI-S systems have been discussed extensively in the literature \cite{Hao:1991,meservey1994spin,tokuyasu1988proximity,tanaka1997theory}. The exchange interaction between the localized magnetic moments of the FI and the itinerant electrons in the SC leads to a  
spin split in the density of states of the latter. Spin split superconductors form an active field of research with varying directions \cite{heikkila2019thermal,bergeret2018colloquium,rouco2019charge,liu2019semiconductor,strambini2017revealing,
ojajarvi2021giant,hijano2022quasiparticle}. 

In this work, we investigate the diode effect in a Josephson junction made of spin-split superconducting electrodes on the 2D surface of a disordered 3D topological insulator (TI).  Specifically, our setup  consists 
of  two spin-split superconductors (FIS) placed on top of a topological insulator, see Fig.  \ref{fig:FISTIFIS}a.
The spin-splitting in the superconductor breaks time-reversal symmetry, whereas inversion symmetry is broken because we consider only the top surface of the topological insulator. Thus, the conditions to have a $\phi_{0}$-effect are fulfilled even without an external magnetic field.
Using the linearized Usadel equation, we first show analytically  that such CPR exhibits the  $\phi_{0}$-effect. Going beyond the linear regime we compute the diode efficiency. Even in the case of low transmission FIS/TI interfaces we obtain an efficiency of 1\%. By increasing  the interface transmission the efficiency can reach  values larger than  7\%. 
We also find that for short junctions the efficiency is maximized at a finite temperature independent of the strength of the exchange field.  

The article is structured as follows. In  section \ref{sec:The system and basic equations} we introduce the  setup and the basic equations. We  derive the  Usadel equation for a  diffusive topological insulator in proximity with a spin-split  superconductor. 
In section \ref{sec:Linearisedcase}, we focus on analytical results that can be obtained by linearizing the Usadel equation, which is valid  under the assumption that the proximity effect is small. Even though in this limit the CPR contains only the first harmonic, and thus has no diode effect, we can determine the condition for observation of the anomalous Josephson currents. The latter is the precursor of a diode effect in the non-linearized equation. We also  show that the $\phi_{0}$- effect is suppressed by impurity  scattering.
In section \ref{sec:Nonlinearisedcase} we go beyond the linear regime and present our numerical results for the nonlinear equation and the diode efficiency as a function of the temperature for different values of the exchange  field, length of the junction, and transparency of the interfaces.  Finally in section \ref{sec:Discussion} we conclude and  propose real material combinations to test our predictions.  Throughout the paper it is assumed that $\hbar = k_{B} = 1$.

\section{The system and basic equations}\label{sec:The system and basic equations}
\begin{figure*}
    \centering
\subcaptionbox{
  }[0.45\linewidth]
  {\hspace*{-2em}\includegraphics[width =8.4cm]{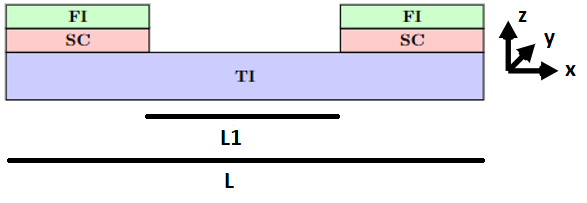}}
\hfill
\subcaptionbox{
  }[0.45\linewidth]
  {\hspace*{-2em}\includegraphics[width =8.4cm]{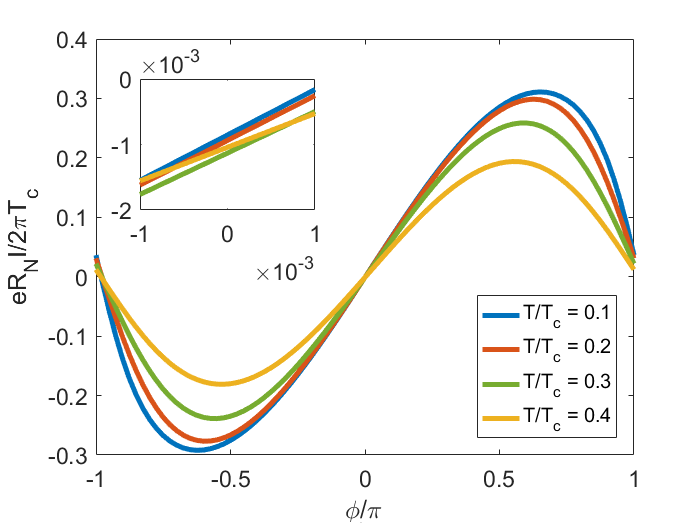}}
    \caption{(a) Sketch of the junction under consideration.  (b) The CPR  of  the FIS-TI-FIS junction for  different temperatures. The parameters chosen are $\frac{\gamma_{0}}{E_{\text{Th}}} = \frac{\Delta_{0}}{E_{\text{Th}}} = \frac{5h}{2E_{\text{Th}}} = 25$, $\frac{L_{1}}{L}=\frac{1}{10}$ and $\frac{l}{L} = 0.08\ll 1$. Inset: zoom in of the CPR around $\phi=0$.
    }
    \label{fig:FISTIFIS}
\end{figure*}

We consider the  FIS-TI-FIS system 
shown in Fig. \ref{fig:FISTIFIS}a. 
  A  FIS can be realized by placing a ferromagnetic insulator  on top of a superconducting film (SC) with a conventional s-wave pair potential. The thickness of the latter is assumed to be small compared to its coherence  length, so that we can assume a homogeneous splitting field in the SC induced by the magnetic proximity effect \cite{hijano2021coexistence}.
We  neglect  the inverse proximity effect of the topological insulator on the superconductor. In this setup a  current can flow through the top surface of the TI from one FIS to the other.    
Because the system is finite in $x$ direction, no current can flow at $x=\pm L/2$, where $L$ is the length of the TI.  We denote by $L_{1}$ the length of the TI between the two FIS electrodes. 

We assume that the transport at the TI  surface is diffusive and can be described by the Usadel equation: the derivation of this equation for our system  closely resembles the derivation of the Usadel equation for a TI in an exchange field as presented in Refs. \cite{zyuzin2016josephson,bobkova2016magnetoelectrics}. However, whereas superconductivity in the systems discussed in these papers is introduced as an effective pair potential, for the spin split superconductor a different approach is taken.

We incorporate the effect of the spin-split superconductor as a self-energy term $\bar{\Sigma}_{s}$. For the self-energy we follow  the approach similar to \cite{tkachov2013suppression,bobkova2017electrically}, in which the self-energy, up to second order in the tunneling parameter $T_{1}$ between the TI and the superconductor, is introduced as
\begin{align}\bar{\Sigma}_{s} = T_{1}^{2}\rho_{S}\check{\tau}_{3}\sigma_{0}\bar{G'}_{S}\sigma_{0}\check{\tau}_{3} = T_{1}^{2}\rho_{S}\bar{G}_{S},
\end{align}
where $\rho_{S}$ is the density of states in the superconductor, $\sigma_{0}$ is the identity matrix in spin space and $\check{\tau}_{3}$ is the third Pauli matrix in electron-hole space. $\bar{G'}_{S}$ is the momentum integrated Green's function in the spin-split superconductor and $\bar{G}_{S} = \check{\tau}_{3}\sigma_{0}\bar{G'}_{S}\sigma_{0}\check{\tau}_{3}$ is introduced to shorten notation. Note that the only effect of this transformation by $\check{\tau}_{3}\sigma_{0}$ is to negate the pair amplitudes.
The self-energy term appears as an added term in the commutator on the left hand side of the Eilenberger equation.
The Eilenberger equation thus reads, using the same presentation in spin-Nambu space as in \cite{zyuzin2016josephson}:
\begin{align}
    &\frac{v_{F}}{2}\{\{\eta_j,\nabla_j \check{g}(1+\vec{n}_{F}\cdot\vec{\eta})\}\} \nonumber\\&= [\check{g}(1+\vec{n}_{F}\cdot\eta),\omega_{n}\check{\tau}_{3}+\Gamma(x)\bar{G}_{S}+\frac{\langle\check{g}(1+\vec{n}_{F}\cdot\vec{\eta})\rangle}{2\tau}],
\end{align}
where  $\{\cdot,\cdot\}$  denotes the anticommutator and $[\cdot,\cdot]$ the commutator. 
 In our notation $\check{g}$ is the quasiclassical Green's function, $\omega_{n}$ is the nth Matsubara frequency, $\vec{n}_{F}$ is the direction of the momentum at the Fermi surface, $v_{F}$ is the magnitude of the Fermi velocity, $\tau$ is the collision time, $\check{\tau}_{3}$ is the third Pauli matrix in Nambu space, $\mu$ is the Fermi energy and $\vec{\eta} = (-\sigma_{2},\sigma_{1},0)$.
  The tunneling parameter $T_{1}$ is nonzero only in FIS regions. To reflect this we introduce the boundary parameter $\Gamma(x)$:
\begin{equation}
    \Gamma(x) = \gamma_{0}\Theta(|x|-\frac{L_{1}}{2})\Theta(\frac{L}{2}-|x|),
\end{equation}
where $\Theta$ denotes the Heaviside function and $\gamma_{0} = T_{1}^{2}\rho_{S}$.

The Green's function is written as $\check{g}\frac{1}{2}(1+\vec{n}_{F}\cdot\vec{\eta})$ to reflect the strong coupling between spin and direction of momentum in a topological insulator.
In this work a 2D surface of a 3D TI is studied. Therefore scattering is not prohibited, unlike in a 1D edge of a 2D TI. We assume the junction is in the dirty limit, that is, the inverse scattering time $\frac{1}{\tau}$ is much larger than any energy scale other than the chemical potential $\mu$. In that case the Green's function $g$ is almost isotropic. Thus, it is a good approximation to keep only the zeroth and first term in the expansion in angular momentum, that is,
\begin{equation}
    \check{g}\approx \check{g}_{s}+\vec{n}_{F}\cdot \vec{\check{g}}_{a},
\end{equation}
where the zeroth order $\check{g}_{s}$ and the first order angular momentum $\vec{\check{g}}_{a}$ satisfy $\check{g}_{s}^2 = \mathbf{1}$ and $\check{g}_{s}\vec{\check{g}}_{a}+\vec{\check{g}}_{a}\check{g}_{s} = \vec{0}$ in order to satisfy the normalisation condition $\bar{g}^{2} = \mathbf{1}$ up to second order in $\tau$. The Green's functions $\check{g}_{s}$ and $\vec{\check{g}}_{a}$ do not have any degrees of freedom in momentum space nor in spin space and are thus functions which map position into the space of 2 by 2 matrices. Using the expansion in angular momentum, the Usadel equation can be derived. The strategy followed to derive the Usadel equation for our structure is very similar to the strategy used in \cite{zyuzin2016josephson}, \cite{bobkova2016magnetoelectrics}. To this end, we first write 
\begin{equation}\label{eq:GS0GS1def}
    \bar{G}_{S} = \check{G}_{S0}\sigma_{0}+\vec{\check{G}}_{S1}\cdot\vec{\sigma},
\end{equation}
where $\check{G}_{S0}$ and $\vec{\check{G}}_{S1}$ are matrix functions without spin degrees of freedom, and $\vec{\sigma}$ is the vector of Pauli matrices in spin space.
It is assumed that the exchange field in the FI always points in the same direction, so that there are no domain walls which may affect the density of states significantly \cite{hijano2022quasiparticle}. The position independent Green's function in the spin-split superconductor is 
\begin{align}
    \bar{G}_{S} &= \frac{1}{2}(1+\vec{b}\cdot\vec{\sigma})\check{G}_{\uparrow}+\frac{1}{2}(1-\vec{b}\cdot\vec{\sigma})\check{G}_{\downarrow}\label{eq:decompositionG}\\
    \check{G}_{\uparrow,\downarrow} &= g_{\uparrow,\downarrow}\tau_{3}+f_{\uparrow,\downarrow}(\cos{\phi}\tau_{1}+\sin\phi\tau_{2})\label{eq:PartsG},
\end{align}
where $g_{\uparrow,\downarrow} = (\omega_{n}\pm ih)/{\sqrt{(\omega_{n}\pm ih)^{2}+\Delta^{2}}}$ are the normal parts and $f_{\uparrow,\downarrow} = \Delta/{\sqrt{(\omega_{n}\pm ih)^{2}+\Delta^{2}}}$ are the anomalous parts \cite{bergeret2005odd}.
The + sign is used for the spin-up component and the - sign for the spin down component, $\tau_{1,2,3}$ are the Pauli matrices in Nambu space,  $h$ is the magnitude of the exchange field $\vec{h} =h\vec{b}$, $\Delta$ is the superconducting potential calculated self-consistently and $\phi$ is the phase of the superconductor. 
Combining  Eq. \eqref{eq:GS0GS1def} with Eqs. \eqref{eq:decompositionG} and \eqref{eq:PartsG}, we may write
\begin{align}
\check{G}_{S0,1} &= g_{S0,1}\tau_{3}+f_{S0,1}(\cos{\phi}\tau_{1}+\sin{\phi}\tau_{2}),\\
\vec{\check{G}}_{S1} &= \check{G}_{S1}\vec{b}\cdot\vec{\sigma},
\end{align}
where $g_{s0,1} = (g_{\uparrow}\pm g_{\downarrow})/2$ and $f_{S0,1} = (f_{\uparrow}\pm f_{\downarrow})/2$.  The component $f_{S0}$ of the condensate is the usual singlet component, whereas $f_{S1}$ is the odd-frequency triplet component \cite{bergeret2005odd}.

The Usadel equation for the angular averaged Green's function $g_{s}$ without spin degrees of freedom in the TI is obtained analogous to the approach laid out in \cite{zyuzin2016josephson}. We combine the spin-trace of the equation obtained by angular averaging and the equation obtained by angular averaging after multiplication by $\vec{n}_{F}$. The resulting Usadel equation is 
\begin{equation}\label{eq:NewUsadel}
    D\hat{\nabla}(\check{g}_{s}\hat{\nabla}\check{g}_{s}) = [\omega_{n} \check{\tau}_{3}+\frac{\Gamma(x)}{2}\check{G}_{S0},\check{g}_{s}],
\end{equation}
with $D = v_{F}^{2}\tau$. Eq. \eqref{eq:NewUsadel} is similar to the Usadel equation in normal metals. However, the derivative is replaced by a generalized derivative:
\begin{align}
    \hat{\nabla} = \nabla+\frac{\Gamma(x)}{2v_{F}}[\cdot,\check{G}_{S1y}]e_{x}-\frac{\Gamma(x)}{2v_{F}}[\cdot,\check{G}_{S1x}]e_{y},
\end{align}
For a spin split superconductor this becomes
\begin{align}\label{eq:DerDef}
    \hat{\nabla} = \nabla+\frac{\Gamma(x)}{2v_{F}}(b_{y}e_{x}-b_{x}e_{y})[\cdot,\check{G}_{S1}].
\end{align}
This derivative is similar to the derivative presented in  \cite{bobkova2016magnetoelectrics} for a topological insulator with an exchange field, in fact, Eq. \eqref{eq:NewUsadel} reduces to this expression if $\check{G}_{S1} = h\check{\tau_{3}}$.
Throughout this work we will assume that the magnetic field is oriented perpendicular to the current direction, so that $b_{y} = 1$ and $b_{x} = 0$.
The equation is accompanied by the boundary conditions
\begin{align}\label{eq:BC}
\hat{\nabla} G_{S1} (x = \pm\frac{L}{2}) = 0.
\end{align}
In this paper it is assumed that the ferromagnetic insulator is either very thin or very thick, so that there is no y nor z-dependence in the problem. As a consequence the equation becomes effectively one-dimensional. 
From the solutions of Eqs. \eqref{eq:NewUsadel},\eqref{eq:DerDef} and \eqref{eq:BC} one can determine the current:
\begin{align}
    I = \frac{\sigma_{N}}{2}T\sum_{n}\text{Tr}\Big(\tau_{3}\bar{G}(x^{*},\omega_{n})\hat{\nabla}\bar{G}(x^{*},\omega_{n})\Big)\; , 
\end{align}
where $\sigma_{N}$ is the normal state conductance, $T$ is the temperature entering the Matsubara frequencies $\omega_{n} = (2n+1)\pi T$,  and $x^{*}$ is any position for which $\Gamma(x^{*}) = 0$.
The quantities of interest in this article are the maximum supercurrents $I_{c}^{+}$ and $|I_{c}^{-}|$ in both directions, and the diode efficiency, defined by
\begin{align}
    \eta = \frac{I_{c}^{+}-|I_{c}^{-}|}{I_{c}^{+}+|I_{c}^{-}|}.
\end{align}
Before showing the numerical solution of the above boundary problem, in the next  section, we  study the linearized equation in the limit of a small proximity effect. As discussed in the introduction,  in this limiting case  the diode effect vanishes, but the anomalous phase $\phi_{0}$ ,  can be studied analytically.  The anomalous current is  a strong indication for the diode effect to appear. 

\section{ Linearized case: The $\phi_{0}$-junction}\label{sec:Linearisedcase}
\begin{figure}
    \centering
    \includegraphics[width = 8.6cm]{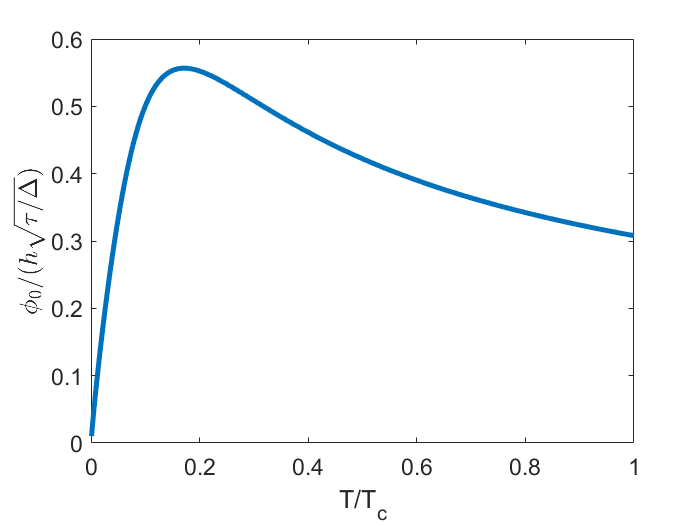}
    \caption{The $\phi_{0}$-effect as a function of temperature as calculated using Eq. \eqref{eq:phi0limit2}. The $\phi_{0}$-effect is suppressed at low temperatures. The $\phi_{0}$-effect is given in units of the small quantity $h\sqrt{\tau/\Delta}$.}
    \label{fig:Phi0linearised}
\end{figure}

To get an understanding  of the physics behind the  new Usadel equation, Eq. \eqref{eq:NewUsadel}, we  focus first on  the case of a weak   proximity effect. 
In this case the anomalous parts of the GF  are much smaller than the normal one, that is, $\text{Tr}(\tau_{1,2}G)\ll\text{Tr}(\tau_{3}G)$, and $G$ can be approximated by  $G(i\omega_{n})\approx\begin{bmatrix}\text{sgn}(\omega_{n})&F\\\Tilde{F}&-\text{sgn}(\omega_{n})\end{bmatrix}$, where $|F|,|\Tilde{F}|\ll 1$.  Using this approximation the Usadel equation reduces to a linear equation. We assume here that  the  system in Fig. \ref{fig:FISTIFIS}a  is infinite in $x$ direction.  The superconductor is only absent in the region $(-\frac{L_{1}}{2},\frac{L_{1}}{2})$ and present everywhere outside this region. In this case Eq. \eqref{eq:NewUsadel} can be written in the three separate spatial regions: 
\begin{align}\label{eq:Usadelperpart}
    \begin{cases}
        D\partial_{xx}F = 2|\omega_{n}|F-\gamma_{0} f_{S0}e^{-i\frac{\phi}{2}}&x<-\frac{L_{1}}{2}\\
        D\partial_{xx}F = 2|\omega_{n}|F&|x|<\frac{L_{1}}{2}\\
    \end{cases}
\end{align}
For $x\xrightarrow{}\pm\infty$ the Green's function should commute with $\omega\tau_{3}+\check{G}_{S0}$.
This implies that the pair potential is given by
\begin{align}
    \lim_{x\xrightarrow{}\pm\infty}F = \frac{\gamma_{0} f_{S0}}{2|\omega_{n}|}e^{\pm i\frac{\phi}{2}}.
\end{align}
These are exactly the same equations as for the conventional SNS junction. However, the equations which joint the solutions at $x = \pm\frac{L_{1}}{2}$, are different compared to the conventional SNS junction. Requiring continuity of both the Green's function and the current through the junction yields
\begin{align}
    &F\left(\pm \frac{L_{1}}{2}+0^{+}\right) =F\left(\pm \frac{L_1}{2}+0^{-}\right)\\
    &\dv{F}{x}\left(\pm \frac{L_{1}}{2}+0^{\pm}\right)+\frac{\gamma_{0}}{v_{F}}f_{S1}\text{sgn}(\omega_{n})e^{\pm i\frac{\phi}{2}}  = \dv{F}{x}\left(\pm \frac{L_{1}}{2}+0^{\mp}\right).\label{eq:CurrentContinuity}
\end{align}
This expression differs from the expression for the SNS junction in the appearance of the $f_{S1}$-term on the right hand side of Eq. \eqref{eq:CurrentContinuity}.
The CPR following from these equations is
\begin{align}
    I(\phi) &= \sum_{n}\frac{1}{4}e^{-2\sqrt{\frac{2|\omega_{n}|}{D}}}\text{Im}((A_{n}-iB_{n})e^{i\frac{\phi}{2}})^{2}\\
    & = \frac{1}{4}\sum_{n}e^{-2\sqrt{\frac{2|\omega_{n}|}{D}}}\left(A_{n}^{2}-B_{n}^{2}\sin{\phi}+2A_{n}B_{n}\cos{\phi}\right)
\end{align}
where $A_{n}$ and $B_{n}$ are real coefficients given by
$A_{n} = \frac{\gamma_{0} f_{S0}}{2|\omega_{n}|}$ and $B_{n} = -i\gamma_{0} f_{S1}\sqrt{\frac{D\omega_{n}}{2}}\frac{1}{|\omega_{n}|}$.
This implies that 
\begin{align}
    \phi_{0} = \arctan2\frac{\sum_{n = -\infty}^{\infty}A_{n}B_{n}e^{-2\sqrt{\frac{2|\omega_{n}|}{D}}}}{\sum_{n = -\infty}^{\infty}(A_{n}^{2}-B_{n}^{2})e^{-2\sqrt{\frac{2|\omega_{n}|}{D}}}}
\end{align}
The $\phi_{0}$-effect is largest if $A_{n} = B_{n}$ and small if $|\frac{B_{n}}{A_{n}}|$ is not close to 1. This ratio is given by
\begin{align}
    |\frac{B_{n}}{A_{n}}| &= \frac{\gamma_{0}  |f_{S1}|\sqrt{\frac{
    D}{2|\omega_{n}|}}2|\omega_{n}|}{\gamma_{0} f_{S0}v_{F}} = \frac{|f_{S1}|}{f_{S0}}\frac{1}{v_{F}}\sqrt{2|\omega_{n}| D}
\end{align}
Recall that the diffusion constant is given by $D = v_{F}^{2}\tau$. This means that $\frac{1}{v_{F}}\sqrt{2|\omega_{n}|D} = O(\sqrt{|\omega_{n}|\tau})$, which is small in the diffusive regime. In fact, in the derivation of the Usadel equation, it is assumed that $\frac{1}{\tau}\gg|\Delta|$ and thus $\frac{1}{\tau}\gg|\omega_{n}|$ for every $n$ that contributes significantly to the current. 
This means the effect can only be large if $f_{S1}\gg f_{S0}$. This constraint can only be satisfied if, $h\gg\sqrt{\omega_{n}^{2}+\Delta^{2}}$ for all Matsubara frequencies that have a significant contribution to the critical current.
However, in the setup discussed in this paper the condition $h\gg\Delta$ can not be realized, since the magnetisation is induced via the superconductor and a high magnetisation destroys the superconductivity. The $\phi_{0}$-effect is suppressed by a factor $\sqrt{|\omega_{n}|\tau}$ in the linearized case.\\
Next the temperature dependence is discussed. Because the $\phi_{0}$-effect is small we can simplify the equation for $\phi_{0}$ to
\begin{align}
    \phi_{0}\approx2\frac{\sum_{n = -\infty}^{\infty}A_{n}B_{n}e^{-2\sqrt{\frac{2|\omega_{n}|}{D}}}}{\sum_{n = -\infty}^{\infty}(A_{n}^{2})e^{-2\sqrt{\frac{2|\omega_{n}|}{D}}}}.
\end{align}
Now, consider the behaviour at low temperatures. 
Since the triplet component $f_{S1}$ is odd in frequency, whereas the singlet component $f_{S0}$ is even in frequency, $|\frac{B_{n}}{A_{n}}|$ is small for small Matsubara frequencies.
As the temperature is decreased these terms become more and more dominant in the sum. Thus, at low temperatures, the $\phi_{0}$-effect is increases with increasing temperature. On the other hand, for large Matsubara frequencies the ratio between triplet and singlet components $|\frac{f_{s1}}{f_{s0}}| = O(\frac{h}{|\omega_{n}|})$. This means that at high temperatures the $\phi_{0}$-effect decreases.
Therefore, the $\phi_{0}$-effect must be non-monotonic as a function of temperature,  it attains a maximum. Moreover, since $\sqrt{\tau}$ is an $\omega_{n}$-independent prefactor it can not determine the maximum. The temperature at which the maximum is attained is determined by only two dimensionless quantities, $\frac{\Delta}{E_{\text{Th}}}$ and $\frac{h}{\Delta}$.\\
An interesting limit is the limit in which $\sqrt{\frac{2\pi T}{D}}L_{1}\ll 1$ and $h\ll\Delta$ so that the exponential suppression can be ignored to first order in $\sqrt{\frac{2\pi T}{D}}L_{1}$. The following expression is obtained \footnote{Strictly speaking, the linearized case leads to a divergency as T goes to 0. We therefore replace $\Delta/|\omega_{n}|$ by $\Delta/\sqrt{\omega^{2}+\gamma_{0}^{2}}$, based on the non-linearized case}:
\begin{align}\label{eq:phi0limit2}
    \phi_{0}\approx&\frac{\sum_{n = 0}^{\infty}A_{n}B_{n}}{\sum_{n = 0}^{\infty}A_{n}^{2}} = h\sqrt{\tau}\frac{\sum_{n = 0}^{\infty}\frac{1}{(\omega_{n}^{2}+\Delta^{2})^{2}\sqrt{\omega_{n}}}}{\sum_{n = 0}^{\infty}\frac{1}{\omega_{n}^{4}+\Delta^{2}\omega_{n}^{2}}}
\end{align}
The multiplication of the sum with $\sqrt{\tau}$ signals the dirty limit suppression of the $\phi_{0}$-effect. The resulting expression is evaluated numerically as a function of temperature.
Numerical evaluation confirmed the non-monotonicity, see Fig. \ref{fig:Phi0linearised}.

In the following sections we discuss the full non-linear equation, and we show that  the diode-effect is non-monotonic with temperature for short junctions. 

\section{Non-linearized case: The superconducting diode effect}\label{sec:Nonlinearisedcase}
To investigate the diode effect one needs to go beyond the linear approach and solve numerically the Usadel equation.  In this section we present our numerical results for the supercurrent in the  FIS-TI-FIS junction. As a first step it is convenient to  write  the  Usadel equation, Eq. \eqref{eq:NewUsadel}, in dimensionless form, normalising $x$ by the total length of the junction. The obtained equation is
\begin{align}
    &\hat{\nabla}(\check{G}\hat{\nabla}\check{G}) = [\frac{\omega_{n}}{E_{\text{Th}}}\check{\tau}_{3}+\frac{\Gamma(x)}{2E_{\text{Th}}}\check{G}_{S0},\check{G}],\\
    &\hat{\nabla} = \dv{}{x}+\frac{\Gamma(x) L}{2v_{F}}\Tilde{b}_{y}[\check{G}_{S1},\cdot]
\end{align}
Thus, all energies are given in units of the Thouless energy $E_{\text{Th}} = \frac{D}{L^{2}}$, whereas lengths are given in units of $L$. 
The strength of the proximity effect is described by the dimensionless parameter $\frac{\gamma_{0} L}{v_{F}}=\frac{\gamma_{0}}{E_{\text{Th}}}\frac{l}{L}$, where $l=v_{F}\tau$ is the mean free path, which in the diffusive limit must be  the shortest length involved in the problem.  
This puts a constraint on the magnitude of the new dimensionless quantity, it must be much smaller than $\frac{\gamma_{0}}{E_{\text{Th}}}$. However, without this term the equations reduce to the equations for the SNS junction, which is known not to have a diode effect \cite{golubov2004current}, it has no time-reversal symmetry breaking. Thus, if the new quantity is very small, the diode effect is very small. Therefore, $\frac{\gamma_{0}}{E_{\text{Th}}}$ must be chosen large, that is, the contact between the superconductor and the TI must be good to have a large $\gamma_{0}$.

To solve the non-linearized Usadel equation, Eq. \eqref{eq:NewUsadel}, the so-called Riccati parametrisation is used,
\begin{equation}
    \check{G} = \frac{1}{1+\Bar{\gamma}\Tilde{\gamma}}\begin{bmatrix}1-\Bar{\gamma}\Tilde{\gamma}&2\gamma\\2\Tilde{\gamma}&-1+\Bar{\gamma}\Tilde{\gamma}\end{bmatrix},
\end{equation}
where $\Bar{\gamma}$ and $\Tilde{\gamma}$ are the Riccati parameters.

In principle,  the pair potential $\Delta$ has to be determined self-consistently since it  is suppressed by  the exchange field \cite{buzdin2005proximity}.   In the numerical calculations   we choose values of the exchange field smaller than  $\frac{h}{\Delta_{0}} = \frac{2}{5}$. Other parameters are set  as follows: $\frac{\gamma_{0}}{E_{\text{Th}}} =  25$, whereas $\frac{L_{1}}{L}$ is chosen to be $\frac{1}{10}$ and $\frac{l}{L} = 0.08\ll 1$. 

 Fig. \ref{fig:FISTIFIS}b shows  our numerical results for the CPR obtained  from the non-linearized  Eq. \eqref{eq:NewUsadel}.  One can see a finite current value at $\phi=0$ associated with the appearance of the anomalous phase  $\phi_{0}$.  Moreover, even though small,  there is a  difference in the absolute value of  the maximum and minimum of the current. 
 This asymmetry reflects the  diode effect. By increasing the temperature   both the current at zero phase and the critical current decrease.

\begin{figure*}
    \centering
\subcaptionbox{
  }[0.45\linewidth]
  {\hspace*{-2em}\includegraphics[width =8.4cm]{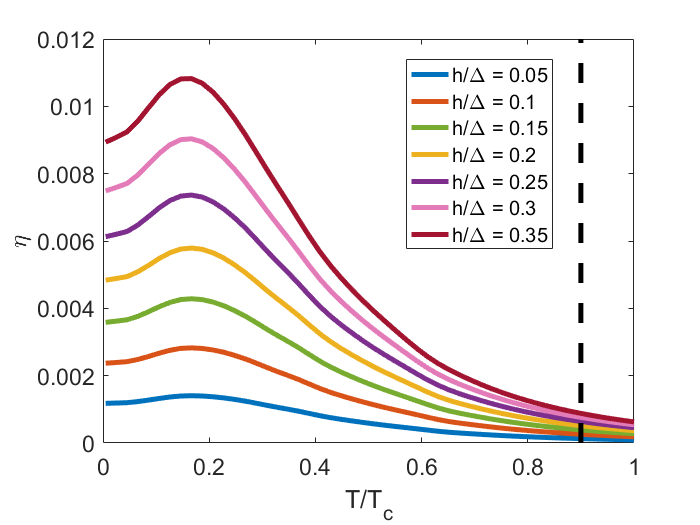}}
\hfill
\subcaptionbox{
  }[0.45\linewidth]
  {\hspace*{-2em}\includegraphics[width =8.4cm]{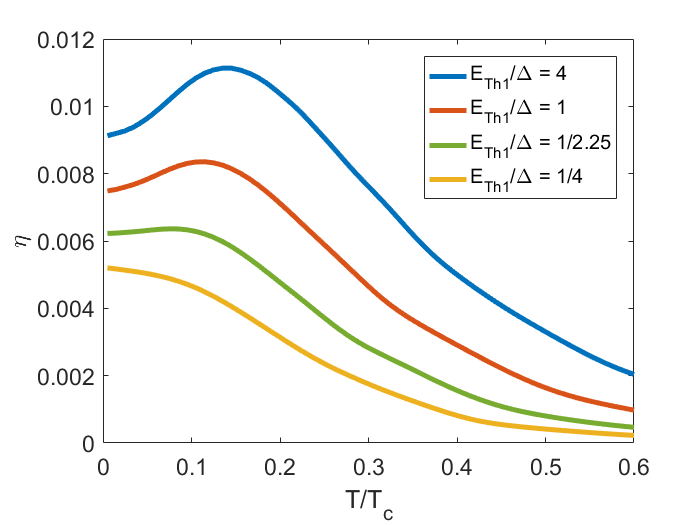}}
    \caption{The temperature dependence of the  diode efficiency  $\eta$ for different values of  the exchange field strength $h$ (a),  and Thouless energy $E_{\text{Th}1} = \frac{D}{L_{1}^{2}}$ of the TI part (b). The critical temperature is for all magnitudes of the exchange field considered here larger than  $0.9T_{c}$, highlighted by the black dotted line.}
    \label{fig:Hdepdiode}
\end{figure*}

We now study the temperature dependence of the diode effect  for different  exchange fields and  sizes of the junction.  The numerical results for the diode efficiency  are shown in Fig. \ref{fig:Hdepdiode}.
 Interestingly, we find a non-monotonic behaviour with a maximum efficiency at a finite temperature, $T_{d}$. 
 It is important to notice, that by computing $\eta$ in Fig. \ref{fig:Hdepdiode}, the self-consistency of the pair potential is ignored, because of the reduction of computational costs. However, we verify that  for all values of $h$ considered here, the self-consistency  leaves the magnitude of the gap almost unchanged for temperatures of the order of  $T = T_{d}$. The
 critical temperatures for all cases shown in Fig. \ref{fig:Hdepdiode}a  lie right of the dashed vertical line. 
 
 If the exchange field is increased,  Fig. \ref{fig:Hdepdiode}a,  the diode efficiency becomes larger.  $\eta$  increases approximately linearly with $h$. The temperature at which the diode efficiency is maximal is almost independent of the exchange field, with $T_{d} \approx 0.18$. 
 
 We also investigate the influence of the distance between the leads, $L_1$  on $\eta$, see Fig. \ref{fig:Hdepdiode}b. 
Specifically, $E_{Th1}=D/L_1^2$  is varied, whereas the quantities $\frac{l}{L}$ and $\frac{L_{1}}{L}$ are held constant.
 As the Thouless energy is decreased, the diode efficiency decreases. Moreover, the temperature $T_{d}$ at which the diode effect is maximal decreases with the length of the junction. For enough long junctions the dependence of $\eta$ on temperature becomes monotonic.

So far,  we have considered  disordered systems with low transparent S/FI interfaces. One can, though, increase  the $\phi_0$ and diode effects by relaxing these conditions.  On the one hand, our  analytical results in section \ref{sec:Linearisedcase} indicate that the $\phi_{0}$-effect can be increased by increasing $\tau$, see for example Eq. (\ref{eq:phi0limit2}). We also verified  numerically   that the diode effect increases if the degree of disorder decreases. 

On the other hand, we also investigated the effect of increasing the ratio $\frac{L_{1}}{L}$. Our  numerical calculations demonstrate  that whereas  $I(\phi = 0)/I_{c}$ increases,  the diode effect decreases with increasing $\frac{L_{1}}{L}$. 
This can be explained as follows. By increasing the  distance $L_1$ between the electrodes  the CPR  becomes more sinus-like. 
To be precise, when increasing $\frac{L_{1}}{L}$ from $\frac{1}{10}$ to $\frac{1}{2}$, the ratio between the magnitudes of the second and first harmonics decreases  from $\approx\frac{1}{6}$ to $\approx\frac{1}{10}$. 
As we discussed before,  
besides the breaking of time-reversal  and inversion symmetries,  the  diode effect relies crucially on the contribution of  higher harmonics to the CPR. 

\begin{figure}
    \centering
    \includegraphics[width = 8.4cm]{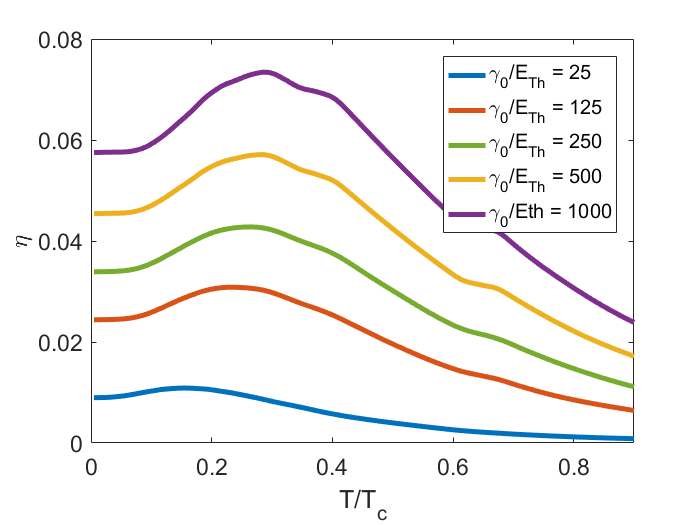}
    \caption{The diode efficiency as a function of temperature for different values of $\gamma_{0}$.}
    \label{fig:GammaDep}
\end{figure}

A way to increase the higher harmonics contribution, is to increase the coupling between the superconducting correlations from the left and right electrodes.  
This can be achieved by increasing the transparency of the FIS/TI interfaces, as shown in Fig. \ref{fig:GammaDep}.

Finally, another way to increase the diode effect is by  increasing  the exchange field, as shown in Fig. \ref{fig:Hdepdiode}. 
In our junction, however, the value of $h$ is limited by the critical field of the superconductor. To increase the strength of the exchange field without suppressing superconductivity in the S electrodes one could add an additional ferromagnetic insulating layers directly on top of the TI between the two superconductors, similar to the situation  investigated in Refs. \cite{zyuzin2016josephson,rabinovich2020electrical,bobkova2020magnetization}.  In that case the exchange field can be larger than the superconducting gap and the  diode effect may increase.

\section{Conclusions}\label{sec:Discussion}

We present a study of the $\phi_0$ and diode effects in a FIS-TI-FIS Josephson junction.  Though disorder tend to suppress them \cite{ilic2021effect}, we found, even in the diffusive limit,   sizable effects without applying any  external field. We found that by  increasing  the FIS/TI interface transparency and the magnetic field  one can increase  the diode effect. For short junctions the diode effect is non-monotonic as a function of temperature. By increasing the distance between the electrodes  the $\phi_{0}$-effect is enhanced, however the diode effect is suppressed  due to the loss of higher harmonics. 

From the point of view of materials the proposed structure can be fabricated with  well studied  material combinations.  On the one  hand the use of   topological insulators in Josephson junctions is well understood \cite{veldhorst2012josephson,veldhorst2013magnetotransport,snelder2014josephson,maier2012induced,sochnikov2015nonsinusoidal,wiedenmann20164pi,oostinga2013josephson,wang20184,mandal2022finite}. On the other hand,  spin-split superconductivity has been measured in several experiments on  ferromagnetic insulator/superconductor bilayers, as for example   EuS/Al structures \cite{moodera1988electron,xiong2011spin,strambini2017revealing,hijano2021coexistence}. Moreover,  good interfaces between  TI and FI has been reported in Ref. \cite{wei2013exchange}.

\section*{Acknowledgements}
We thank Stefan Ilic for useful discussions. 
We acknowledge  financial support from Spanish AEI through project PID2020-114252GB-I00 (SPIRIT).
FSB acknowledges financial support from  the European Union’s Horizon 2020 Research and Innovation Framework Programme under Grant No. 800923 (SUPERTED), the A. v. Humboldt Foundation, and  the Basque Government through grant  IT-1591-22. 

\bibliography{apssamp.bib}
\end{document}